\documentclass[utf8]{frontiersinFPHY_FAMS}

\usepackage[utf8]{inputenc}
\usepackage[T1]{fontenc}
\usepackage{anyfontsize} 
\usepackage{url,hyperref,lineno,microtype,subcaption}
\usepackage[onehalfspacing]{setspace}
\usepackage{mathrsfs}

\setcitestyle{square} 
\newtheorem{remark}{Remark}

\newtheorem{proposition}{Proposition}

\def\keyFont{\fontsize{8}{11}\helveticabold }
\def\firstAuthorLast{Olascoaga \& Beron-Vera} \def\Authors{M.J.\
Olascoaga\,$^{1,*}$ and F.J.\ Beron-Vera\,$^{2}$}
\def\Address{$^{1}$Department of Ocean Sciences, Rosenstiel Shool
of Marine, Atmospheric, and Earth Science, University of Miami, FL,
USA\\ $^{2}$Department of Atmospheric Sciences, Rosenstiel Shool
of Marine, Atmospheric, and Earth Science, University of Miami, FL,
USA} \def\corrAuthor{Corresponding Author}
\def\corrEmail{jolascoaga@miami.edu}

\def\pct{\%{ }}
\newcommand{\defn}[1]{\emph{#1}}

\def\X{X}
\def\t{t}
\def\D{\mathcal D}
\def\A{\mathcal A}
\def\B{\mathcal B}
\def\S{\mathcal S}
\def\O{\mathcal O}
\def\R{\mathcal R}
\def\P{\mathcal P}
\def\1{\mathbf 1}
\def\q{\mathbf q}

\def\v{\mathbf v}
\def\o{\mathbf o}
\def\k{\mathbf k}
\def\p{\boldsymbol\pi}
\def\Id{\ensuremath{\operatorname{Id}}}
\def\pct{\%}

\begin{document}
\onecolumn
\firstpage{1}

\title[TPT-based oil spill prediction]{Exploring the use of
Transition Path Theory in building an oil spill prediction scheme}

\author[\firstAuthorLast]{\Authors}
\address{}
\correspondance{}
\extraAuth{}
\maketitle

\begin{abstract}
  The Transition Path Theory (TPT) of complex systems has proven a
  robust means for statistically characterizing the ensemble of
  trajectories that connect any two preset flow regions, say $\mathcal
  A$ and $\B$, directly.  More specifically, transition paths are
  such that they start in $\A$ and then go to $\B$ without detouring
  back to $\A$ or $\mathcal B$. This way, they make an effective
  contribution to the transport from $\A$ to $\B$.  Here, we explore
  its use for building a scheme that enables predicting the evolution
  of an oil spill in the ocean.  This involves appropriately adapting
  TPT such that it includes a reservoir that pumps oil into a
  typically open domain. Additionally, we lift up the restriction
  of the oil not to return to the spill site en route to a region
  that there is interest to be protected. TPT is applied on oil
  trajectories available up to the present, e.g., as integrated
  using velocities produced by a data assimilative system or as
  inferred from high-frequency radars, to make a prediction of
  transition oil paths beyond, without relying on forecasted oil
  trajectories.  As a proof of concept we consider a hypothetical
  oil spill in the Trion oil field, under development within the
  Perdido Foldbelt in the northwestern Gulf of Mexico, and the
  \emph{Deepwater Horizon} oil spill.  This is done using trajectories
  integrated from climatological and hindcast surface velocity and
  winds as well as produced by satellite-tracked surface drifting
  buoys, in each case discretized into a Markov chain that provides
  a framework for the TPT-based prediction.
  
  \keyFont{\section{Keywords:} oil spill, Transition Path Theory,
  Markov chain}
\end{abstract}

\section{Introduction}

The \defn{Transition Path Theory} \cite{E-vandenEijnden-06,
E-vandenEijnden-10} provides a rigorous statistical means for
highlighting the dominant pathways connecting two metastable states
of a complex dynamical system.  That is, instead of studying the
individual complicated paths connecting them, TPT concerns their
average behavior and shows their dominant transition channels.  TPT
produces a much cleaner picture, and hence much easier to interpret,
allowing one to frame the relative contribution of myriad competing
paths in the presence of stochasticity.

Applications of TPT have now grown beyond the study of molecular
systems \cite{Noe-etal-09, Metzner-etal-09, Meng-etal-16, Thiede-etal-19,
Liu-etal-19, Strahan-etal-21}, for which it was originally conceived.
Such applications involve reactions, which, to develop fully, must
overcome barriers in the energy landscape.  Indeed, TPT has been
recently used to investigate tipping atmospheric phenomena such as
sudden stratospheric warmings \cite{Finkel-etal-20, Finkel-etal-21}.
But departing from the rare event framing setting is possible and
motivated by the fact that in other type of applications, particularly
fluid mechanics, there is a basic interest of understanding how two
regions (of the flow domain) are most effectively connected.  With
this idea in mind, TPT has been used in oceanography to bring
additional insight to pollution \cite{Miron-etal-21-Chaos} and
macroalgae pathways in the ocean \cite{Beron-etal-22-Chaos}, as
well as paths of the upper \cite{Drouin-etal-22} and lower
\cite{Miron-etal-22, Beron-etal-22-JPO} limbs of the meridional
overturning circulation in the Atlantic Ocean.

In this paper, we explore the use of TPT in building an oil spill
prediction scheme that relies on oil parcel trajectory information
up to the time of the prediction. That is, the scheme assumes that
time-resolved validated model velocities are available up to the
present time when the prediction is made.  No assumption is made
on the availability of forecasted velocities from numerical models.
The basic assumption is that environmental conditions, namely,
near-surface ocean currents and winds, prior to the prediction
instant prevail for some time past this instant.  The scheme,
however, enables updating the predictions over time as new velocity
information becomes available.  This idea was put forth in
\cite{Olascoaga-Haller-12} to predict sudden changes in the shape
of the oil slick from the \emph{Deepwater Horizon} spill, yet using
an approach different than the that one proposed here based on TPT.
The framework for TPT is a Markov chain on boxes resulting by
discretizing of the oil parcel motion. Such a framework has been
used in \cite{Perez-etal-21}, but to describe general oil spill
scenarios based on climatological velocities.

The rest of the paper is organized as follows.  In Section \ref{sec:TPT}
we review the formulae of TPT for autonomous, i.e., time-homogeneous,
discrete-time Markov chains.  Section \ref{sec:TPT_oil} presents
an extension of the standard TPT setup for the case of oil spills.
The TPT-based prediction scheme is presented in Section
\ref{sec:TPT_prediction}.  In Section \ref{sec:res} we test the
scheme assuming a hypothetical spill in the northwestern Gulf of
Mexico (Section \ref{sec:res_trion}) and by considering the
\emph{Deepwater Horizon} oil spill (Section \ref{sec:res_dwh}).
Finally, in Section \ref{sec:conclusions} we present a summary of
the paper along with several ideas as to how improving the proposed
prediction scheme.
 
\section{Methods}

\subsection{Transition Path Theory for Markov
chains}\label{sec:TPT}

Let $\X_n$ denote the position of a random walker at discrete time
$nT$, $n \in \mathbb Z$, $T>0$, in a \emph{closed} two-dimensional
domain $\D$ covered by $N$ disjoint boxes $\{B_1,\dotsc,B_N\}$.
(For simplicity, we will avoid using a different notation for the
covering of $\D$.  Also, $\X_n\in B_i$ and $i:B_i\in\D$ (or similar)
will be simplified to $\X_n = i$ and $i\in\D$, respectively.) Then
$\Pr(\X_{n+1} = j) = \sum_{i\in\D} P_{ij}\Pr(\X_n = i)$ where
\begin{equation}
  P_{ij} := \Pr(\X_{n+1} = j \mid \X_n = i),\quad
  \sum_{j\in\D}P_{ij} = 1\,\forall i\in\D,
  \label{eq:P}
\end{equation}
which describes the proportion of probability mass in $B_i$ that
flows to $B_j$ during $T$, called a \emph{transition time step}.
The row-stochastic matrix $\smash{P = (P_{ij})_{i,j\in\D}}$ is
called the \defn{transition matrix} of the (two-sided) autonomous,
discrete-time Markov chain $\{\X_n\}_{n\in \mathbb{Z}}$.  We assume
that the process is ergodic and mixing with respect to the
\defn{stationary distribution} $\p = (\pi_i)_{i\in\D}$.  Namely, a
componentwise positive vector on $\D$, \emph{seen as an $N$-dimensional
state (vector) space}, which is invariant and limiting.  Normalized
to a probability vector, i.e., such that $\smash{\sum_{i\in\D}}\pi_i
= 1$, $\p$ satisfies $\p = \p P = \v P^\infty$ for any probability
vector $\v$ (on $\D$).  For details, cf.\ \cite{Norris-98}.

The \defn{Transition Path Theory} (\defn{TPT}) provides a rigorous
characterization of the ensemble of trajectory pieces, which,
\emph{flowing out last from a region $\A \subset \D$, next go to a
region $\B \subset \D$, disconnected from $\A$}.  Such trajectory
pieces are called \defn{reactive trajectories}.  This terminology
\cite{E-vandenEijnden-06, E-vandenEijnden-10} originates in chemistry,
where $\A$ (resp., $\B$) is identified with the reactant (resp.,
product) of a chemical transformation.  The fluidic interpretation
of reactive trajectories is of trajectories of \emph{diffusive
tracers} that contribute to the bulk transport between $\A$ and
$\B$, which can be thought as a \defn{source} and a \defn{sink} or
\defn{target}, respectively.

\begin{remark}
  For a diffusive tracer to fit the above interpretation, its
  evolution must be described by a stationary stochastic process,
  i.e., an advection--diffusion equation with a steady velocity.
  This can be seen by discretizing its Lagrangian motion using, for
  instance, Ulam’s method (e.g., \cite{Kovacs-Tel-89, Koltai-10}),
  which consists in projecting the probability density of finding
  tracer on a given spatial location at a discrete time instant
  onto a finite-dimensional vector space spanned by indicator
  functions on boxes, which, covering the flow domain, are normalized
  by their Lebesgue measure (area) \cite{Miron-etal-19-Chaos}. The
  boxes represent the states of the autonomous, discrete-time Markov
  chain that the diffusive tracer parcels wander about.
\end{remark}

The main objects of TPT are the \emph{forward}, $\smash{\q^+ =
(q^+_i)_{i\in\D}}$, and \emph{backward}, $\smash{\q^- = (q^-_i)_{i\in\D}}$,
\defn{committor probabilities}.  These give the probability of a
trajectory initially in $B_i$ to first enter $\B$ and last exit
$\A$, respectively.  Namely, 
\begin{equation}
  q_i^\pm : = \Pr(\t^\pm_{\B} < \t^\pm_{\A} \mid  \X_0 = i)
\end{equation}
where 
\begin{equation}
  \t^\pm_{\S} := \pm\inf\{nT \ge 0 : \X^\pm_n\in\S\},\quad 
  \inf\emptyset := \infty,
\end{equation}
with the plus (resp., minus) sign denoting (random) \emph{first
entrance} (resp., \emph{last exit}) \emph{time} of a set $\S\subset\D$.
Here, $\{\X^-_n\}_{n\in \mathbb{Z}}$ is the original chain,
$\{\X_{n}\}_{n\in \mathbb{Z}} = \{\X^+_n\}_{n\in \mathbb{Z}}$, but
traversed in backward time, i.e., $\X^-_n := \X_{-n}$. The reversed
chain's transition matrix, $P^- = (P^-_{ij})_{i,j\in\D}$, where
\begin{equation}
  P^-_{ij} := \Pr(\X^-_{n+1} = j\mid\X^-_n = i) = \Pr(\X_n = j\mid\X_{n+1} = i) =
  \frac{\pi_j}{\pi_i}P_{ji},
\end{equation}
since $\Pr(\X_n = i) = \pi_i$.  The committors are fully
determined by $P$ and $\boldsymbol\pi$ by solving two linear
algebraic systems with appropriate boundary conditions \cite{Metzner-etal-09,
Helfmann-etal-20}, concisely written as:
\begin{equation}
  q^\pm_i = \sum_{j\in\D} P^\pm_{ij}q^\pm_j,\,i\in\D\setminus\!\A\cup\B,
  \quad q^\pm_{i\in\mathcal A} = \delta_{1\mp1,2},\quad q^\pm_{i\in\B}
  = \delta_{1\pm1,2},  
  \label{eq:q}
\end{equation}	
where $P^+_{ij} = P_{ij}$ and $\delta_{kl}$ is Kronecker's delta.

Four main statistics of the ensemble of reactive trajectories are
expressed using the committor probabilities as follows:
\begin{enumerate}
 \item The \defn{reactive probability distribution},
	$\smash{\boldsymbol \pi^{\A\B} = (\pi^{\A\B}_i)_{i\in\D}}$,
	where $\pi^{\A\B}_i$ is defined as the joint probability
	that a trajectory is in box $B_i$ while transitioning from
	$\A$ to $\B$.  This is computed as \cite{Metzner-etal-09,
	Helfmann-etal-20}
   \begin{equation}
    \pi^{\A\B}_i = q^-_i\pi_iq^+_i.
	 \label{eq:piAB}
   \end{equation} 
 \item The \defn{reactive probability flux}, $\smash{f^+ =
   (f^+_{ij})_{i,j\in\D}}$, where $f^+_{ij}$ gives the \emph{net}
   flux of trajectories going through $B_i$ and $B_j$ in one time
   step on their direct way from $\A$ to $\B$, indicates the dominant
   transition channels from $\A$ to $\B$. According to \cite{Noe-etal-09,
   Helfmann-etal-20}, this is computed as:
   \begin{equation}
	f^+_{ij} := \max\left\{f^{\A\B}_{ij} -
   f^{\A\B}_{ji},0\right\},\quad
	f^{\A\B}_{ij} = q^-_i \pi_i P_{ij}q^+_j.
	\end{equation}
 \item The \defn{reactive rate} of trajectories leaving
	$\A$ or entering $\B$, defined as the probability per time
	step of a reactive trajectory to leave $\A$ or enter $\B$,
	is computed as \cite{Metzner-etal-09, Helfmann-etal-20}
   \begin{equation}
	k^{\A\B} := \sum_{i:\in\A,j\in \D}
	f^{\A\B}_{ij} \equiv \sum_{i\in\D,j\in\B} f^{\A\B}_{ij}.
   \end{equation}
	Divided by the transition time step $T$, $k^{\A\B}$ is
	interpreted as the \emph{frequency} of transition paths
	leaving $\A$ or entering $\B$ \cite{Miron-etal-21-Chaos}.
 \item Finally, the \defn{reaction duration}, $t^{\A\mathcal
   B}$, of a transition from $\A$ to $\B$ is obtained by dividing
   the probability of being reactive by the reactive rate,
	interpreted as a frequency
   \cite{Helfmann-etal-20}:
   \begin{equation}
    t^{\A\B} := \frac{\sum_{i\in\D}\pi^{\A\B}_i}{k^{\A\B}/T}.
	 \label{eq:tAB}
   \end{equation}
\end{enumerate}

\subsection{Adapting Transition Path Theory to the oil spill
problem}\label{sec:TPT_oil}

Let $x(t)$ represent a very long \emph{oil} parcel trajectory
visiting every box of the covering of $\D$. Assume that this not
different than any other trajectory, namely, it represents a
realization of a stationary random process. Then $x(t)$ and $x(t+T)$
at any $t>0$ provide observations for $\X_n$ and $\X_{n+1}$,
respectively.  Under these conditions, we can approximate $P_{ij}$
via counting transitions between covering boxes, viz.,
\begin{equation}
  P_{ij} \approx \frac{\#\{x(t)\in B_i,\, x(t+T)\in
  B_j\}}{\#\{x(t) \in B_i,\}},\, t:\text{any}.  
  \label{eq:P_estimation}
\end{equation}

In general the domain $\mathcal D$ potentially affected by an oil
spill will represent some portion of the ocean.  This makes $\D$
an \emph{open} flow domain.  In such a case, $P$ cannot be
row-stochastic, which requires an adaptation of TPT. To achieve the
required adaptation, we first replace $P$ by a row-stochastic
transition matrix $\tilde P$ defined by
\begin{equation}
  \tilde P := 
  \begin{pmatrix}
    P & \mathbf p^{\D\to\omega} & \mathbf 0\\ 
	 \mathbf p^{\D\gets\omega}
    & 0 & 0\\ 
	 \mathbf p^{\O\gets\R} & 0 & p^{\R\to \R}
  \end{pmatrix}
  \label{eq:P_closed}
\end{equation}
on the extended $(N+2)$-dimensional state space 
\begin{equation}
  \tilde\D := \D\cup\omega\cup\R
  \label{eq:D_closed}
\end{equation}
Here, $\omega$ denotes a virtual state, called a \defn{two-way
nirvana state}, which absorbs probability mass imbalance from $\D$
and sends it back to the chain.  More precisely, in \eqref{eq:P_closed},
\begin{equation}
  \mathbf p^{\D\to\omega} = \Big(1 - \sum\nolimits_{j\in \D}
  P_{ij}\Big)_{i\in\D}
\end{equation}
gives the outflow from $\mathcal D$, while $\mathbf p^{\D\gets\omega}$
gives the inflow, which can be constructed, as we do below, using
reentry information available from trajectory data outside $\D$.
In turn, $\R$ is another virtual state, called an \defn{oil reservoir
state}, from which the chain drains probability mass through the
oil spill site $\O\subset\D$.  That is,
\begin{equation}
  \mathbf p^{\O\gets\R} = \frac{\1_\O}{|\O|}\big(1 - p^{\R\to
  \R}\big),
\end{equation}
so that $\sum_{i\in\D} p_i^{\O\gets\R} + p^{\R\to \R} = 1$. (The
notation $\1_\S$ is used to mean a vector on (the $N$-dimensional
space given by the covering of) $\D$ with ones in the entries
corresponding to (subcovering) $\S\subset\D$ and zeros elsewhere.)
Below we will expand on how to set $p^{\R\to \R}$.  As in the
standard TPT setup presented in Section \ref{sec:TPT}, the stochastic
process described by $\tilde P$ in \eqref{eq:P_closed} is assumed
to be in stationarity.  The stationary distribution on $\tilde\D$
is denoted $\tilde\p$.  (Herein a tilde is used to emphasize that
the quantity in question is computed using the extended Markov chain
on $\tilde\D$.)  A caveat to note is that the Markov process on
$\tilde\D$ cannot be strictly ergodic because $\R$ is never visited
by a trajectory unless it starts there.  Yet, this does not rule
out the existence of a well-defined $\tilde\p$ (unless $p^{\R\to\R}
= 0$, in which case $\tilde\p\vert_\R = 0$, and hence $\tilde\p$
will not be strictly componentwise positive).

Now, arguably, it is the oil that reaches the coastline or any
region one may want to protect what really matters, irrespective
that oil trajectories visit the spill site many times in between.
Call this region $\P\subset\D$. The former cannot be achieved by
simply setting $\A = \O$ and $\B = \P$.
\begin{proposition}
  To achieve the desired effect, which slightly deviates us from
  the standard TPT setting, one must set $\A = \R\cup\omega$ and
  $\B = \P$ for the computation of $\tilde\q^+$ for extended chain
  on $\tilde\D$, while $\A = \R$ and $\B = \P\cup\omega$ for the
  computation of $\tilde\q^-$ (on $\tilde\D$).  
  \label{pr:reservoir}
\end{proposition}

\begin{figure}[t!]
  \centering%
  \includegraphics[width=.5\textwidth]{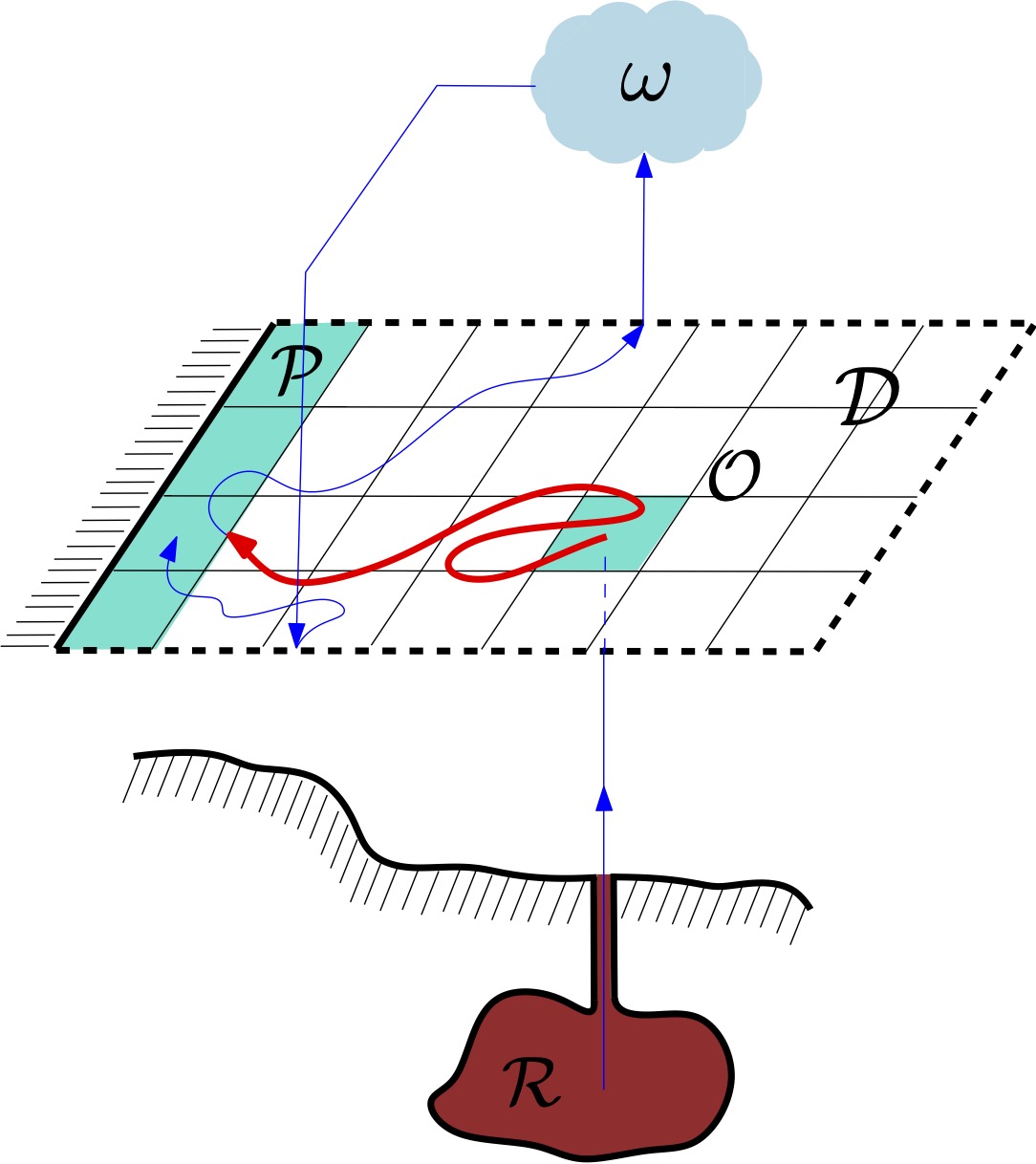}%
  \caption{The framework for the TPT-based prediction scheme is an
  autonomous, discrete-time Markov chain on a state space given by
  the box covering of a two-dimensional, open ocean domain, $\D$,
  augmented by two virtual boxes or states. One state, called a
  two-way nirvana state and denoted by $\omega$, compensates for
  probability mass imbalances due to the openness of $\D$.  The
  other state, called an oil reservoir state and denoted by $\R$,
  injects probability mass into the chain through the spill site,
  $\O\subset\D$.  Highlighted in red is the restriction to $\D$ of
  a reactive trajectory connecting $\R$ with $\P\subset\D$, a region
  that there is interest in being protected, chosen to be the
  shoreline in the cartoon. Such a trajectory flows last from $\R$
  and next goes to $\P$, while being constrained to stay in $\D$,
  once it enters $\D$. So defined, a reactive trajectory of oil may
  return back many times to the spill site before reaching the
  protected area. TPT provides a statistical characterization of
  the ensemble of reactive trajectories, highlighting the
  dominant paths of oil into $\P$.}
  \label{fig:tpt-cartoon}%
\end{figure}

Indeed, placing the source in $\R$ enables oil paths visiting $\O$,
and including $\omega$ as indicated prevents trajectories from
escaping the flow domain, thereby highlighting the portion that
flows into $\P$ through $\D$ (Fig.\@~\ref{fig:tpt-cartoon}).  The
TPT formulae in Section \ref{sec:TPT} remains the same with the
above choices of $\A\subset\tilde\D$ and $\B\subset\tilde\D$, and,
of course, the use of $\tilde P$ and $\tilde\p$ in place of $P$ and
$\p$, respectively, in them.

Inclusion of a two-way-nirvana state is not new, as it was first
applied by \cite{Miron-etal-21-Chaos} to treat transition paths of
marine debris into the subtropical oceans great garbage patches.
Additional TPT applications involving this type of closure include
\cite{Miron-etal-22, Beron-etal-22-Chaos, Beron-etal-22-JPO,
Drouin-etal-22}.  The use of an oil reservoir state is novel in
TPT.

\subsection{A proposal for using Transition Path Theory to
predict oil paths and arrival}\label{sec:TPT_prediction}

We propose to apply TPT such that it makes use of available oil
trajectories up to the present, to make a prediction beyond, so it
does not rely on forecasted oil trajectories.  This follows precedent
work \cite{Olascoaga-Haller-12}, which was able to predict sudden
changes in the shape of the oil slick during the \emph{Deepwater
Horizon} spill.  The expectation is that such a type of prediction
should be superior than that based on forecasted velocities, which
are not validated by data as is the case of hindcast velocities
form an analysis system.  Moreover, when available, the scheme
enables the use of velocities inferred from high-frequency radar
measurements and even trajectories of appropriate satellite-tracked
surface drifting buoys.

The proposed prediction scheme more specifically consists in applying
TPT using trajectories over a few time steps prior to the prediction
time, say $t_0 = n_0T$. That is, we propose to compute the closed
transition matrix $\tilde P$ for the augmented Markov chain on
$\tilde\D$, as given by \eqref{eq:P_closed}, and the various TPT
statistics from it according to Proposition \ref{pr:reservoir}, by
making use of \emph{all trajectories available over $t \in \{(n_0-m)T,
\dotsc, (n_0-1)T, n_0T\}$ for some $m \le n_0$}. This way a prediction
for the spilled oil distribution on $t = (n_0+1)T$, \emph{in direct
transition into the region to be protected}, $\P$, is obtained.
The skeleton thereof will be provided by the \emph{two}-dimensional
vector field taking values at discrete positions $x_i$, where $x_i$
is the center of box $B_i\in\D$, given by
\begin{equation}
  J(x_i) := \sum_{j\in\D}\tilde f^+_{ij} e(i,j),
  \label{eq:J}
\end{equation}
where $e(i,j)$ is the (two-dimensional) unit vector pointing from
$x_i$ to $x_j$, the center of box $B_j\in\D$. The above is a visualization
means of reactive probability flux, proposed in \cite{Helfmann-etal-20}.
We will refer to \eqref{eq:J} as the \defn{reactive current} at
position $x_i$.  The prediction can be updated, as we will do in
the examples we provide below, by computing TPT using trajectories
within time windows \emph{sliding} over the prediction time $t_0$
(or $n_0$).

In the present exploratory work, the oil evolution in every case
will be obtained by pushing forward a probability vector on $\tilde\D$
with support in $\R$ at time $t = 0$, namely,
\begin{equation}
  \tilde \o(0) = \frac{\tilde \1_\R}{|\R|},
\end{equation}
under right multiplication by a \emph{nonautonomous} version of the
augmented chain transition matrix $\tilde P$ in \eqref{eq:P_closed}.
Denoted by $\tilde P(n)$ to make explicit its dependence on $t =
nT$, this will be constructed by \emph{accounting for the start
time $t$ in the estimation of $P$ in} \eqref{eq:P_estimation}. To
add a bit of extra realism, we will set $p^{\R\to \R}$ in
\eqref{eq:P_closed} to
\begin{equation}
  p^{\R\to \R}(n) = \frac{1}{N-n+1},
  \label{eq:pRR}
\end{equation}
representing an oil reservoir that is drying over the time interval
$t \in \{0, T, 2T, \dotsc, NT\}$.  Indeed, when $n = N$, $p^{\R\to
\R} = 1$, meaning that the probability mass flux into the ocean
domain $\D$ through the oil spill site $\O$ is nil. That is, the
reservoir $\R$ has completely dried out by $t = NT$. This assumption
may be adapted based on any information available about how quickly
the oil reservoir may be expected to empty or a spilling oil well
may be capped.  The \emph{total accumulated} oil, which flows out
from $\R$ into $\D$ through $\O$ in an uninterruptedly but decaying
in time manner, at discrete time $t = nT$ will be given by $\o(n)
= \tilde\o(n)\vert_\D$, where
\begin{equation}
  \tilde\o(n)  = \tilde\o(0) \left(\Id +
  \sum_{k=0}^{n-1}\prod_{l=k}^{n-1}\tilde P(l)\right).  
  \label{eq:oil}
\end{equation}
Note that $\tilde\o(n)$, and hence $\o(n)$, does not need to be a
probability vector, and the units in which $\o(n)$ is measured are
determined by the units assigned to $\tilde\o(0)$.

To incorporate the effects of a drying oil reservoir in the TPT
prediction step, the autonomous $\tilde P$ used in that step will
have to set $p^{\R\to \R}$ in \eqref{eq:P_closed} to the average
value of \label{eq:pRR} over the corresponding time interval.

The idea of using trajectory information up to the present prevents
us from using the extension of TPT for time-inhomogeneous Markov
chains proposed in \cite{Helfmann-etal-20}, as it might be thought
to be more suitable for a prediction scheme in a naturally time-varying
environment.  The reason is that, as formulated, nonautonomous TPT
requires unavailable trajectory information and knowledge of when
$\P$ will be hit by transition paths.

A final comment is reserved to the oil trajectories themselves.  If
$u(x,t)$ denotes the surface ocean velocity, as output from an ocean
circulation model, and $u_{10}(x,t)$ is the wind velocity at 10-m
above the sea surface, as produced by some atmospheric circulation
model, the oil trajectories will here be obtained by integrating
\begin{equation}
  \dot x = u(x,t) + \alpha u_{10}(x,t),\quad 0 < \alpha \ll 1,
  \label{eq:dxdt}
\end{equation}
for many initial conditions over a domain including the ocean domain
$\D$ of interest, so reentry information, namely, that required to
evaluate $\mathbf p^{\O\gets\omega}$ in \eqref{eq:P_closed}, is
available.  Equation \eqref{eq:dxdt} is a \emph{minimal} law for
oil parcel motion \cite{Abascal-etal-09, Abascal-etal-12}.  It
exclusively accounts for the wind action on oil parcels, neglecting
the various physical and biogechemical changes oil undergo as it
spends time in the environment, collectively known as ``weathering''
effects \cite{Spaulding-17}.  Typically employed values of $\alpha$
range from 2 to 4\pct{ }\cite{ASCE-96}.

\section{Results}\label{sec:res}

\subsection{Hypothetical oil spill in the Trion
field}\label{sec:res_trion}

We begin by applying our proposed TPT-based prediction scheme to a
hypothetical oil spill in the Trion field, located within the Perdido
Foldbelt, a geological formation in the northeastern Gulf of Mexico
with an important oil reservoir for ultradeepwater drilling under
development \cite{Trion-20}.  This will be done by considering two
different velocity representations.

In the first representation, $u$ in \eqref{eq:dxdt} is chosen to
be given by a daily climatology of surface velocity constructed
from velocities over 18-yr (1995--2012) produced by a free-running
regional configuration for the Gulf of Mexico at $1/36^{\circ}$-horizontal
resolution \cite{Jouanno-etal-16} of the ocean component of the
Nucleus for European Modelling of the Ocean (NEMO) system
\cite{Madec-16}. This dataset was used in \cite{Gough-etal-19} to
investigate persistent passive tracer transport patterns using
so-called climatological Lagrangian Coherent Structures (cLCSs)
\cite{Duran-etal-18}.  A main finding was the presence of a mesoscale
hook-like cLCs providing a barrier for cross-shelf transport nearly
year round.  Consistent with the motion of historical satellite-tracked
drifting buoys, with the majority of them including a drogue, albeit
shallow (cf.\ \citet{Miron-etal-17} for details), synthetic drifters
originating beyond the shelf were found to be initially attracted
to this cLCS as they spread anticyclonically and eventually over
the deep ocean.  In \cite{Gough-etal-19} it is noted that this
should have implications for the mitigation of contaminant accidents
such as oil spills.  This picture, however, may be altered for oil,
as this is expected to be influenced by the wind action, which in
\cite{Gough-etal-19} was not accounted for, when winds are strong.
To evaluate their effect we need a representation for $u_{10}$ in
\eqref{eq:dxdt}, which is chosen to be provided by daily climatological
wind velocity at 10-m height from the European Centre
for Medium-Range Weather Forecasts (ECMWF) atmospheric reanalysis
ERA-Interim \cite{Dee-etal-11}.

\begin{figure}[t!]
  \centering%
  \includegraphics[width=\textwidth]{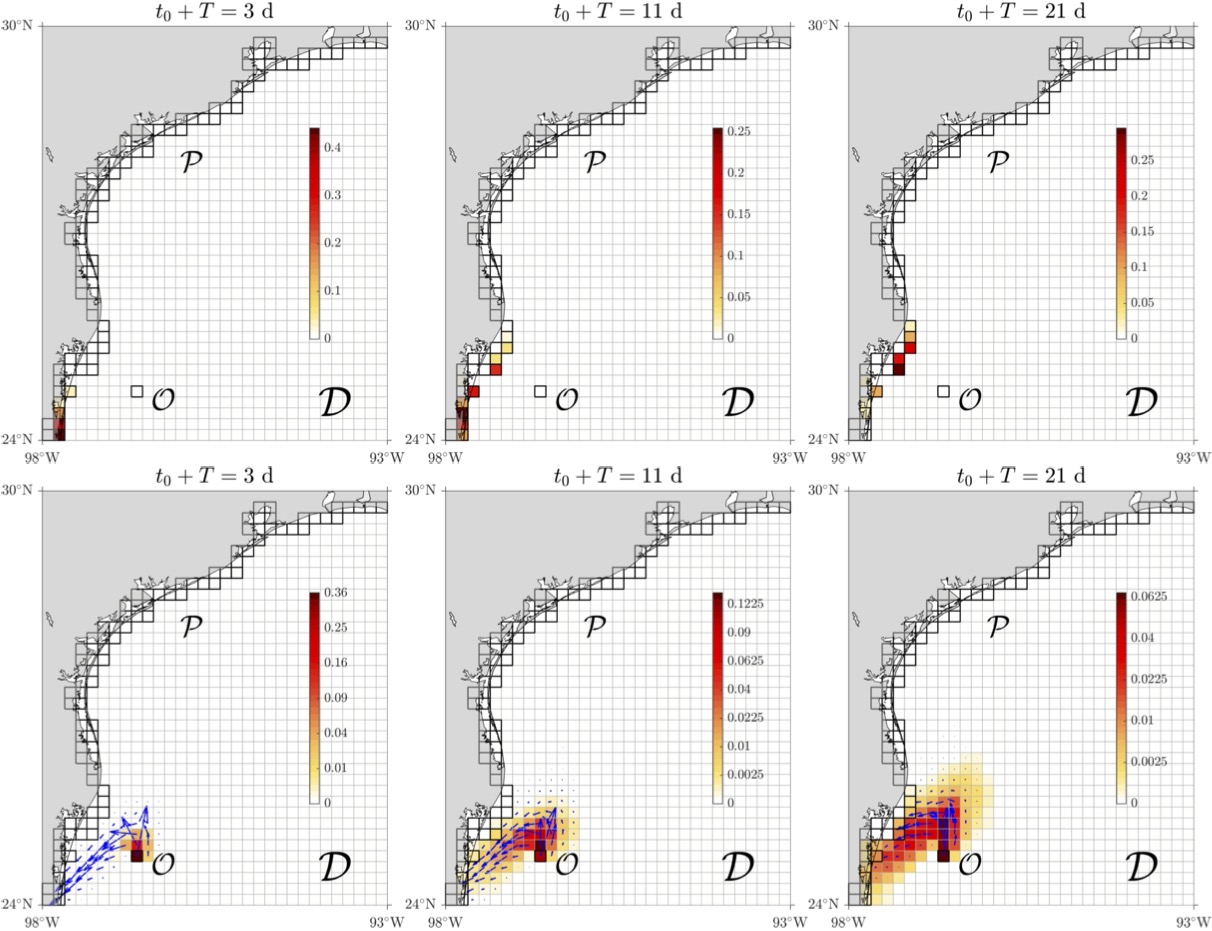}%
  \caption{(top panels) Based on the daily climatological NEMO$+$ECMW
  oil velocity model, predictions on time $t = t_0$ of (normalized)
  reactive rates of arrival of transition paths, through the open
  northwestern Gulf of Mexico domain $\D$ and into the coastline
  ($\P$), of oil emerging from a hypothetical open well in the Trion
  field ($\O$), located within the Perdido Foldbelt, on $t = 0$,
  corresponding to February 1. The transition time step, $T = 1$\,d.
  (bottom panels) Relative distribution of accumulated oil on $t =
  t_0 + T$, overlaid with predicted (on $t=t_0$) reactive currents,
  indicating transition oil paths into $\P$.}
  \label{fig:tpt-nemo}%
\end{figure}

In Fig.\@~\ref{fig:tpt-nemo} we present our first set of results.
These are based on the use of trajectories obtained by numerically
integrating \eqref{eq:dxdt}, with the daily climatological NEMO$+$ECMW
velocity data above, using a 4th-order Runge--Kutta scheme with
cubic interpolation in space and time.  The integrations, reinitialized
every day along the month of February, span $T = 1$\,d. We consider
initial conditions distributed uniformly over an ocean domain larger
(by $1^{\circ}$ to the east and south) than that one ($\D$) contained
inside [$98^{\circ}$W,\,$93^{\circ}$W] $\times$
[24$^{\circ}$S,\,30$^{\circ}$N], shown in Fig.\@~\ref{fig:tpt-nemo}.
To evaluate the transition matrix on $\D$, using \eqref{eq:P_estimation},
we cover $\D$ with boxes of about $1/6^{\circ}$-side, including
roughly 100 test points per box when trajectories initialized once
are only considered.  The transition time step $T = 1$\,d guarantees
sufficient loss of memory into the past for the Markovian assumption
to hold; indeed, the typical decorrelation timescale on ocean surface
is not longer than 1\,d as estimated from drogued drifting buoys
\cite{LaCasce-08}, and is likely to be even shorter when the wind
action is accounted for.  Stationary of the Markov chain on the
extended domain $\tilde\D$ is checked numerically.  That is, we
check that the largest eigenvalue of the transition matrix $\tilde
P$ has multiplicity 1, and is equal to 1, to numerical precision.
However, in the computation of $\P$, namely, the transition matrix
on the open domain $\D$, we make sure to allow as much communication
as possible along the corresponding Markov chain by applying applying
Tarjan's \cite{Tarjan-72} algorithm on the associated directed
graph, as we have done earlier work (e.g., \cite{Miron-etal-17,
Miron-etal-19-JPO, Miron-etal-19-Chaos, Beron-etal-20-Chaos}).  This
can result in some boxes of the covering of the ocean domain to be
excluded, particularly when dealing with observed (satellite-tracked)
trajectories, as we consider in Section \ref{sec:res_dwh}, below.

With $\O$, the oil spill site, taken to be a box of the covering
of $\D$ closest to the Trion field and $\P$, the region to be
protected, taken to be the coastal boxes, the top row of
Fig.\@~\ref{fig:tpt-nemo} shows the \emph{predicted} reactive rate
of (oil) trajectories entering each box of $\P$ for selected times
since since 1 February when the oil spill is hypothetically initiated.
Specifically, we show this normalized to a probability vector on
$\D$, viz.,
\begin{equation}
  \k := \left.\sum_{i\in\P}\frac{\tilde k^{\R B_i}}{\sum_{i\in\P}
  \tilde k^{\R B_i}}\1_{B_i}\right\vert_\P,\quad \tilde k^{\R B_i} =
  \sum_{l\in\tilde\D}
  \tilde f_{li} = \sum_{l\in\tilde\D} \tilde f_{il},\,
  i\in\P, 
  \label{eq:kP}
\end{equation}
where the TPT computation follows Proposition \ref{pr:reservoir}.
More precisely, in the TPT calculation we apply Proposition
\ref{pr:reservoir} \emph{for each $i\in\P$}, i.e., with $\P$ replaced
by $i\in\P$. The transition matrix of the augmented chain in
$\tilde\D$, given by \eqref{eq:P_closed}, is computed using
trajectories over $t\in \{t_0-2T,t_0-T,t_0\}$ for every $t_0$ in
Fig.\@~\ref{fig:tpt-nemo}.  The coastal boxes that are predicted
to be most affected by the oil spill correspond are those where
components of $\k$ in \eqref{eq:kP} take the largest values.  Note,
in this case, that the predicted coastal boxes to be most affected
change over time, moving from the boxes corresponding to the Mexican
state of Tamaulipas north, toward the international border with the
United States.  This is consistent with the updated reactive current
predictions and, most importantly, with the portion of the simulated
spilled oil distribution directed into the coastline.  This is
depicted in the bottom row of Fig.\@~\ref{fig:tpt-nemo}.  More
specifically, the heatmap in each panel is of $\o(n_0+1) =
\tilde\o(n_0+1)\vert_\D$, with $\tilde\o(n)$ given in \eqref{eq:oil},
but normalized to a probability vector, giving the \emph{relative
distribution} of oil that has accumulated on $\D$ at time $t = t_0+T
= (n_0+1)T$.  Overlaid on each heatmap are the predicted reactive
currents \eqref{eq:J} on each $t_0$.  These are seen to anticipate,
$T = 1$-d in advance, the motion of the oil directed into the
coastline quite well.

An important observation is that the main responsible for this
motion is the wind action on the oil, which makes it to bypass the
cross-shelf transport barrier for passive tracers shown
\cite{Gough-etal-19} to be supported nearly year-round by the
climatological NEMO surface ocean velocity field. Indeed, the winter
season is dominated by ``nortes'' \cite{Gomez-Resendiz-02}.  These
are strong, predominantly northerly winds suddenly produced after
the passage of a cold fronts, which, imprinted in the daily
climatology, promote the accumulation of the oil toward the coastline.

In addition to predicting the transition paths of oil the into the
coastline, TPT can give a prediction for the arrival time of the
oil.  Using the reaction duration formula \eqref{eq:tAB}, but on
the extended Markov chain on $\tilde\D$ and for each $i\in\P$, as
done to compute the reaction rate \eqref{eq:kP}, we compute on $t_0
= 2$\,d, the first time a prediction can be made, that
\begin{equation}
  \min_{i\in\P} \big\{\tilde t^{\R B_i}\big\} =  \min_{i\in\P}
  \left\{\frac{\sum_{l\in\tilde\D}\tilde\pi_l^{\R B_i}}{\tilde
  k^{\R B_i}/T}\right\} 
  \approx 14\,\text{d}.
  \label{eq:tP}
\end{equation}
This early prediction turns out be somewhat longer than the actual
arrival time to the coast, which happens around 11-d after the
(simulated) oil started.  This assessment is rough, based on when
oil probability mass is found for the first in a coastal box,
independent of how much.  With this in mind, early prediction
\eqref{eq:tP} is not that off at all.  But one may wonder if it
could be updated with time, i.e., as newer data becomes available.
This is hopeless using \eqref{eq:tAB}, as it computes the duration
of the whole reaction from source to target, which are fixed in
space.  However, the desired update of the arrival time prediction
may indeed be accomplished.  We discuss how in the last section.

\begin{figure}[t!]
  \centering%
  \includegraphics[width=\textwidth]{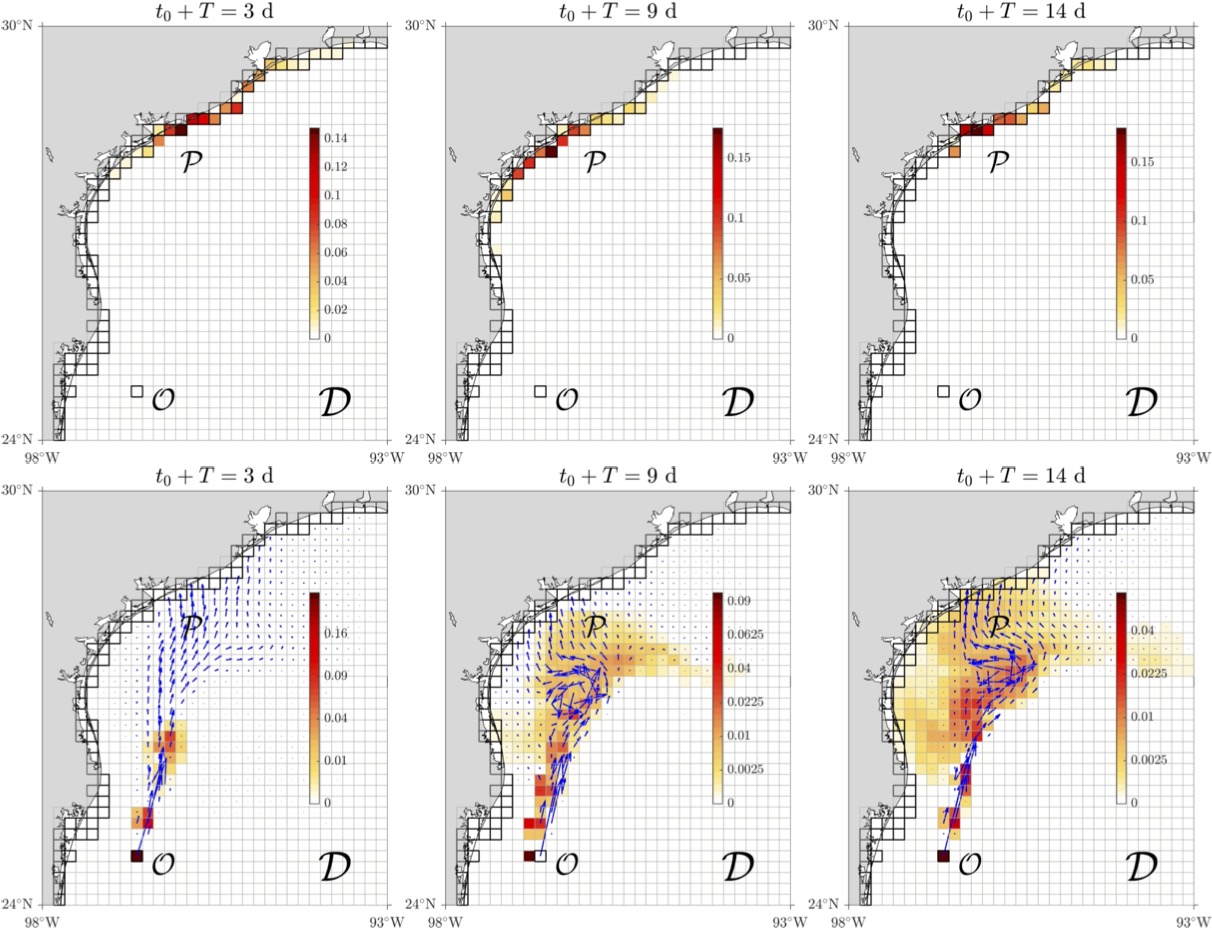}%
  \caption{As in Fig.\@~\ref{fig:tpt-nemo}, but based on the
  NEMO$+$ECMWF oil velocity model and $t = 0$ corresponding to
  22 July 2022.}
  \label{fig:tpt-hycom}%
\end{figure}

The second representation for $u$ in \eqref{eq:dxdt} that we consider
is provided by hourly surface ocean velocity output from the HYCOM
(HYbrid-Coordinate Ocean Model) $+$ NCODA (Navy Coupled Ocean Data
Assimilation) Gulf of Mexico $1/25^{\circ}$ Analysis
(GOMu0.04/expt\_90.1m000) \cite{Chassignet-etal-07}.  For $u_{10}$
in \eqref{eq:dxdt} we use three-hourly wind velocity, 10-m above
the sea surface, from the National Centers for Environmental
Prediction (NCEP) operational Global Forecast System (GFS) analysis
and forecast at $1/4^{\circ}$-horizontal resolution \cite{NCEP-15}.
In neither case we considered forecasted velocities, but a record,
from 22 July 2022 through 8 August 2022, of \emph{hincast} velocities,
i.e., as produced by the systems while they assimilated observations
``on the fly'' to make the forecasts.  The TPT setup for the
HYCOM$+$NCEP oil velocity is similar to that for the climatological
NEMO$+$ECMWF oil velocity.  For instance, trajectory integrations
are reinitialized daily and span $T = 1$\,d, and the number of boxes
of the domain partition is similar with a comparable number of test
points per box.  Unlike the climatological case, the time origin
($t = 0$) of the oil spill simulation corresponds to a specific day
of the current year, chosen to be 22 July 2022.  The simulation
extends out to 8 August 2022.  Covering a summer time period, it
is not affected by ``nortes'' wind events, which prevail in winter.
The results are shown in Fig.\@~\ref{fig:tpt-hycom}.  As can be
expected, an important difference with those shown in
Fig.\@~\ref{fig:tpt-nemo} is a stronger influence of the cross-shelf
transport barrier for passive tracers on the distribution of the
simulated oil, which, while eventually reaching the coastline, by
$t = 16$\,d it starts to develop a hook-like shape pointing into
the open ocean, similar to that described in \cite{Gough-etal-19}.
This happens after part of the oil is trapped in an anticyclonic
circulation.  The predicted arrival location on $t_0 = 2$ d falls
quite close to the arrival location, which takes place between the
southern Texas cities of Corpus Christi and Galveston on $t \approx
9$\,d.  The early prediction (on $t_0 = 2$\,d) of arrival time is
$t \approx 6$\,d, which shorter than the arrival time, calling for
an update.

Overall the above results provide support to our proposed TPT-based
oil spill prediction scheme, based on the assumption that the motion
prior to the prediction time is representative of motion beyond it,
for some time, which we test against observations in the section
that follows.

\subsection{The \emph{Deepwater Horizon} oil
spill}\label{sec:res_dwh}

\begin{figure}[h!]
  \centering%
  \includegraphics[width=\textwidth]{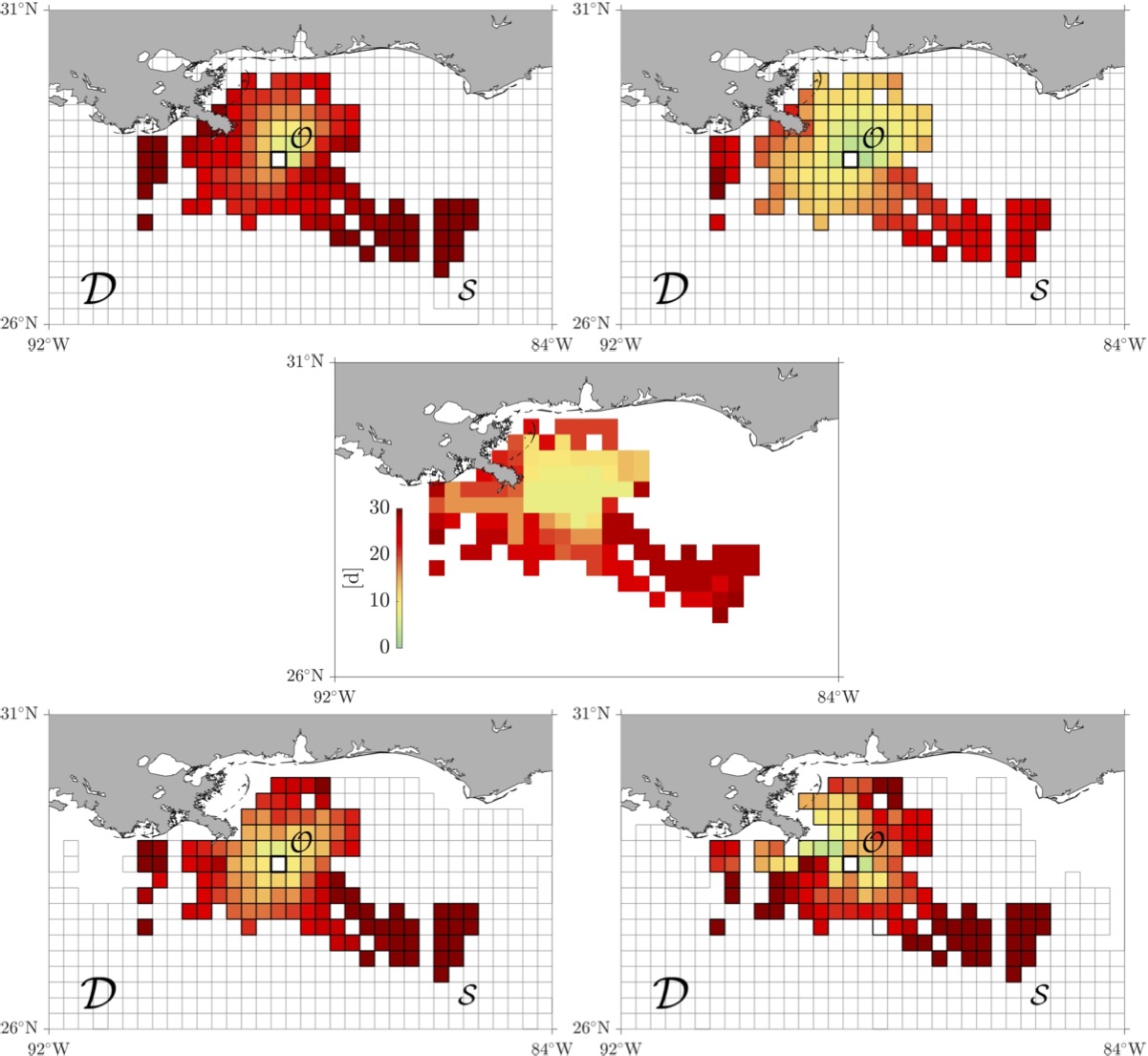}%
  \caption{Estimated from satellite images of the \emph{Deepwater
  Horizon} oil slick, time to first find oil on the surface of the
  ocean since the 22 April 2010, when the spill started to spill
  from the Macondo well (center panel), along with reaction duration
  of transition paths into each box of the subset $\S$ of the domain
  covering $\D$ that most closely intersects the region visited by
  the surfaced oil according to: trajectories integrated from surface
  NCOM velocities (top-left panel) and the latter with the addition
  of $3\pct$ NOAA/NCEI windage (top-right panel) over a period of
  7\,d prior the first time oil was found on the ocean surface on
  29 April 2010; and trajectories produced by historical satellite-tracked
  surface drifters with drogue present (bottom-left panel) and
  absent (bottom-right panel).}
  \label{fig:dwh}%
\end{figure}

The basic assumption on which the TPT-based oil prediction scheme
builds on, namely, that environmental conditions prior to the
prediction time can be prolonged beyond it, for some time, is here
tested using oil arrival time estimates for the \emph{Deepwater
Horizon} spill \cite{Crone-Tolstoy-10}.  Shown in the heatmap in
the center panel of Fig.\@~\ref{fig:dwh}, the arrival time estimates
are inferred using available satellite images of the oil slick, as
produced by the National Oceanic and Atmospheric Administration
(NOAA) National Environmental Satellite Data and Information Service
(NESDIS) Marine Pollution Surveillance Program \citep{Street-11}.
The time origin is 22 April 2010, when the Macondo well started to
spill oil due to the sinking of the mobile offshore rig after an
explosion caused by a blowout two days before \cite{Crone-Tolstoy-10}.
The value assigned to each colored pixel corresponds to the first
time (in d) the oil visited that pixel.

Four TPT-based estimates, based on four different Markov chains,
of the arrival time are shown in Fig.\@~ \ref{fig:dwh}, two in the
top panels and two in bottom panels.  More specifically, these are
reaction durations \eqref{eq:tAB} into each box ($B_i$) of the set
$\S$ of the domain covering ($\D$) that most closely intersects the
region in the center panel of Fig.\@~\ref{fig:dwh} where the oil
was observed to be occupied in the satellite imagery. More specifically,
we show
\begin{equation}
  \mathbf t := \left.\sum_{i\in\S} \tilde t^{\R B_i}
  \1_{B_i}\right\vert_\S,
\end{equation}
computed using Proposition \ref{pr:reservoir} with $\P$ replaced
by $i\in\S$. The Markov chains are constructed as follows. 

For the top panels of  Fig.\@~\ref{fig:dwh}, we use $T=1$\,d-long
trajectories integrated from \eqref{eq:dxdt} in the interval 22
April 2010 through 29 April 2010, the day when the satellite images
record reveal oil on the surface for the first time (since the spill
started, on 22 April 2010). For both panels we use $u$ in \eqref{eq:dxdt}
represented by daily surface ocean velocities produced by the
experimental real-time Intra-Americas Sea Nowcast/Forecast System
(IASNF) at $1/25^{\circ}$-horizontal resolution, which is based on
the U.S.\ Naval Coast Ocean Model (NCOM) \cite{Ko-etal-03}.  In
turn, the wind ($u_{10}$) velocity representation is obtained from
daily $1/4^{\circ}$-horizontal resolution winds at 10\,m from the
NOAA/National Centers for Environmental Information (NCEI) Blended
Sea Winds product \cite{Zhang-etal-06}.  The difference between the
top-left and right panels is that in the former the oil model
velocity uses $\alpha = 0$, i.e., the wind effect is shut off, and
in the latter $\alpha = 0.03$, as we have set above.  The size of
each box of the covering is of $1/25^{\circ} \times 1/25^{\circ}$,
and the number of test points per box is (roughly) $100$.

For the bottom panels of Fig.\@~\ref{fig:dwh}, we consider pieces
of length $T = 1$d of historical, i.e., available since 1992 to
date, satellite-tracked trajectories of surface drifting buoys from
the NOAA Global Drifter Program (GDP) \cite{Lumpkin-Pazos-07} with
the following distinction: in the left panel, we consider drifters
that have their (15-m-long) drogues (sea anchors) attached at all
times, while in the right panel, only those that do note have a
drogued attached during the extent of the record (because this has
been lost at the beginning of the record, after deployment, as
assessed by the drogue presence algorithm of \cite{Lumpkin-etal-12},
or because the drifter was intentionally deployed undrogued).  The
size of each box of the covering is of $1/25^{\circ} \times
1/25^{\circ}$ for the drogued and the undrogued drifters.  The
number of test points per box is small compared to that of the
simulated oil trajectories, with only about $10$ test points per
box.

Comparison of the top-right panel of Fig.\@~\ref{fig:dwh} with the
center panel reveals that our assumption on that simulated trajectories
up to the prediction time can be used to indeed make reliable
predictions (beyond) holds quite well, at least for environmental
conditions prevalent during the \emph{Deepwater Horizon} spill, and
during the timespan covered by the satellite imagery of the oil
slick.  In this case, windage in the minimal oil parcel trajectory
model \eqref{eq:dxdt} does not dramatically impact the TPT-based
computation, as can be seen from the comparison of the top-right
panel with top-left panel. Recall that these use trajectories
integrated over 7-d prior to first time oil is observed on the
ocean surface and that the satellite oil images record extends for
30\,d.  Moreover, even TPT results based on historical drifter
trajectories are reasonable, irrespective of whether the drifters
are drogued or undrogued, as it follows from the inspection of the
bottom panels of Fig.\@~\ref{fig:dwh}.  These results might not
come as a big surprise, as an analysis of daily climatological model
velocities were successful in reproducing the ``tiger tail'' shape
produced by the \emph{Deepwater Horizon} oil slick \cite{Duran-etal-18}.
Similarly, the analysis of altimetry-derived surface ocean velocities
was capable of reproducing a similar shape into which drifting buoys
from the Grand Lagrangian Deployment (GLAD) organized along
\cite{Olascoaga-etal-13}.

Clearly, the above results for the \emph{Deepwater Horizon} oil
spill, the largest and best document oil spill, may not be made
extensible to other oil spills in other regions of the ocean and
year seasons with more variable environmental conditions reigning.
Yet, they are an encouraging sign of the validity of the assumption
on which our TPT-based oil prediction scheme builds on.  In such
more variable environments, a more sophisticated model than
\eqref{eq:dxdt} may be needed and there is also ample space to
improving the TPT setup.  We highlight possible or required
improvements below.

\section{Summary and concluding remarks}\label{sec:conclusions}

In this paper, we have given the first steps toward building an oil
spill prediction scheme based on the use of Transition Path Theory
(TPT) for autonomous, discrete-time Markov chains on boxes, which
cover a typically open flow domain, an result under an appropriate
discretization of the oil motion, assumed to be described by a
stationary stochastic process, namely, to obey an advection--diffusion
with a steady velocity.  Transition paths highlight the main conduits
of communication between a source and a target in the phase of a
dynamical system under noise, and thus they can be used to unveil
the main routes of oil from an accidental spill in the ocean into
a region that needs protection.

The basic premise on which the TPT-based oil prediction scheme is
that trajectory information up to the prediction time can be used
to infer oil motion beyond it.  The TPT setup deviates from the
standard TPT setup in that one needs to cope with the openness of
ocean flow domain where a spill takes place, which is accounted for
by the addition of a virtual box (state) that compensates for
probability mass imbalances, and also with a way to represent the
injection of oil in to the open flow domain, which is done via the
addition of another virtual state representing an oil reservoir.
The scheme was tested by considering a hypothetical oil spill in
the Trion field, located within the Perdido Foldbelt in the
northwestern Gulf of Mexico, and the \emph{Deepwater Horizon} oil
spill, giving good signs of its validity.

Several improvements to the proposed scheme are possible or required.
These should help increasing the quality of the predictions,
particularly under environmental conditions more variable than those
of the situations considered here.
\begin{itemize}
  \item In the examples considered, the prediction time increment,
  say $\Delta t_0$, was chosen to be equal to the transition time
  step ($T$).  For the Markovianity assumption to be fulfilled, $T$
  should not be taken shorter than 1\,d, the typical Lagrangian
  decorrelation time in the surface ocean.  However, there is no
  restriction on the choice of $\Delta t_0$ and the frequency of
  prediction updates may be higher than daily.
  \item In a similar manner as the prediction of transition paths
  of oil into the region one desires to protect are updated over
  time, the duration of the paths should be updated, as it is not
  just where oil will end that is important to know, but when it
  will arrive at the protected area.  This will require one to
  compute the reaction duration into the target region from any
  place in between it and the source. Mathematically, this is given
  by the expectation of the random time to first enter the target
  conditioned on starting on any box (state) of the chain while the
  trajectory is reactive. A collection of such boxes can be chosen
  to be those where the updated reaction distribution \eqref{eq:piAB},
  which tells one where the reactive flux bottlenecks, acquires the
  largest values. There currently is no TPT formula that accounts
  for this in the case of the Markov chain setting of this paper,
  but one can be derived by combining potential theory with a
  Sisyphus chain, which agrees with the original chain when this
  is reactive and is mapped to a nirvana state when is not [Luzie
  Helfmann, 2022, private communication].  For diffusion processes,
  a related statistic is derived in \cite{Finkel-etal-21}.
  \item Another aspect that we have not accounted for is oil beaching.
  This is an additional source of openness of the flow domain.
  Beaching has been incorporated in a physically consistent manner
  in the problem of plastic pollution \cite{Miron-etal-21-Chaos}.
  Such a solution does not seem appropriate for the oil problem,
  and beaching may necessary result in a nonstationary Markov chain.
  The nonautonomous extension of TPT in \cite{Helfmann-etal-20}
  does not require the Markov chain to be in stationary, which may
  provide a resolution to this aspect.  However, nonautonmous TPT
  requires trajectory information past the prediction time, and
  thus a different strategy to cope with beaching will need to be
  designed.
  \item Last but not least is the oil parcel trajectory model.  We
  have considered the minimal possible model, which only accounts
  for windage in a bulk manner.  The typically used $3\pct$ windage
  accounts for the effects wave-induced Stokes drift, which may be
  explicitly added to the ocean surface velocity with the corresponding
  reduction of the windage.  The Stokes drift may be obtained from
  a wave model.  The full ocean surface plus wave-induced drift is
  measured, partially at least \cite{Graber-etal-97, Rohrs-etal-15},
  by high-frequency radars, which, when available, may me be easily
  incorporated.  Additional improvements may be provided by the
  Maxey--Riley theory for floating material on the ocean surface
  \cite{Beron-etal-19-PoF, Olascoaga-etal-20}, which includes a law
  for windage depending on buoyancy in closed form, or consideration
  of the output from an oil spill trajectory model like OpenDrift
  \cite{Dagestad-etal-18}, which accounts for weathering effects
  (e.g., \cite{Perez-etal-21, Romo-etal-22}).  The trajectories
  produced by the minimal model or improvements thereof may be
  combined with trajectories of satellite-tracked with appropriate
  drifting buoys, if these are deployed in the area where the oil
  spill takes place.
\end{itemize}

\section*{Author contributions}

The authors equally contributed to the paper.

\section*{Funding}

Support for this work was provided by Consejo Nacional de Ciencia
y Tecnolog\'{\i}a (CONACyT)--Secretar\'{\i}a de Energ\'{\i}a (SENER)
as part of the Consorcio de Investigaci\'on del Golfo de M\'exico
(CIGoM).

\section*{Acknowledgments}

We thank Luzie Helfmann and P\'eter Koltai for discussions on
adapting TPT to the oil spill problem, and Joaquin Tri\~nanes
for making available to us the \emph{Deepwater Horizon} oil
spill imagery.


\section*{Data availability statement}

The ECMWF/ERA Interim wind renanalysis can be obtained from
\href{https://apps.ecmwf.int/datasets/data/interim-full-daily/levtype=sfc/}{https://\allowbreak
apps.ecmwf.int/\allowbreak datasets/data/\allowbreak
interim-full-daily/\allowbreak levtype=sfc/}. The Gulf of Mexico
HYCOM$+$NCODA analysis data are available at
\href{https://www.hycom.org/data/gomu0pt04/expt-90pt1m000}{https://\allowbreak
www.hycom.org/\allowbreak data/\allowbreak gomu0pt04/\allowbreak
expt-90pt1m000}. The NCEP/GFS wind data can be retrieved from
\href{https://www.ncei.noaa.gov/products/weather-climate-models/global-forecast}{https://\allowbreak
www.ncei.noaa.gov/\allowbreak products/\allowbreak
weather-climate-models/\allowbreak global-forecast}. Digitized
versions of the observed surface ocean oil distribution images
during the \emph{Deepwater Horizon} spill are available at
\href{https://satepsanone.nesdis.noaa.gov/pub/OMS/disasters/DeepwaterHorizon/composites/2010/}{https://\allowbreak
satepsanone.nesdis.\allowbreak  noaa.gov/pub/\allowbreak
OMS/disasters/\allowbreak DeepwaterHorizon/\allowbreak composites/2010/}.
The NCOM/IASNFS output is available from
\href{https://www.northerngulfinstitute.org/edac/oceanNomads/IASNFS.php}{https://\allowbreak
www.\allowbreak northerngulfinstitute.org/\allowbreak
edac/oceanNomads/\allowbreak IASNFS.php}. The NOAA/NEIC wind data
are available from
\href{https://coastwatch.pfeg.noaa.gov/erddap/griddap/ncdcOwClm9505.html}{https://\allowbreak
coastwatch.pfeg.\allowbreak noaa.gov/\allowbreak erddap/griddap/\allowbreak
ncdcOwClm9505.html}. The NOAA/GDP drifter data are available at
\href{https://www.aoml.noaa.gov/phod/gdp/}{https://\allowbreak
www.aoml.\allowbreak noaa.gov/\allowbreak phod/\allowbreak gdp/}.

\bibliographystyle{Frontiers-Vancouver}

\end{document}